\documentclass{article}
\usepackage{spconf,amsmath,graphicx}

\usepackage{amsfonts}
\usepackage{here}
\usepackage{amssymb}
\usepackage{enumerate}
\usepackage{graphicx}
\usepackage{multirow}
\usepackage{tabularx}
\usepackage{arydshln}
\usepackage{color}
\usepackage{float}
\usepackage{bm}
\usepackage{comment}
\usepackage{cite}
\usepackage{amsmath}
\usepackage{algorithmic}
\usepackage{array}
\usepackage{setspace}
\usepackage{subfigure}
\usepackage{colortbl}
\usepackage{url}

\newcommand{\bi}[1]{\ensuremath{\boldsymbol{#1}}}   

\newlength\savedwidth
\newcommand{\wcline}[1]{\noalign{\global\savedwidth\arrayrulewidth\global\arrayrulewidth 1.0pt} \cline{#1}
\noalign{\global\arrayrulewidth\savedwidth}}

\usepackage{tabularx}
\newcolumntype{Y}{&gt;{\centering\arraybackslash}X} 

\newcommand{\figenv}{\def\@captype{figure}}
\newcommand{\tabenv}{\def\@captype{table}}
\usepackage{}

\title{Joint Analysis of Sound Events and Acoustic Scenes\\Using Multitask Learning}
\name{Noriyuki Tonami$^{\dagger}$, Keisuke Imoto$^{\dagger}$$^\ddagger$, Ryosuke Yamanishi$^{\dagger}$, Yoichi Yamashita$^{\dagger}$
}
\address{$^{\dagger}$\hspace{1pt}Ritsumeikan University, $^{\ddagger}$\hspace{1pt}Doshisha University}

\begin{document}
\maketitle
\begin{abstract}
%
Sound event detection (SED) and acoustic scene classification (ASC) are important research topics in environmental sound analysis.
Many research groups have addressed SED and ASC using neural-network-based methods, such as the convolutional neural network (CNN), recurrent neural network (RNN), and convolutional recurrent neural network (CRNN).
The conventional methods address SED and ASC separately even though sound events and acoustic scenes are closely related to each other.
For example, in the acoustic scene ``office,'' the sound events ``mouse clicking'' and ``keyboard typing'' are likely to occur.
Therefore, it is expected that information on sound events and acoustic scenes will be of mutual aid for SED and ASC.
In this paper, we propose multitask learning for joint analysis of sound events and acoustic scenes, in which the parts of the networks holding information on sound events and acoustic scenes in common are shared.
Experimental results obtained using the TUT Sound Events 2016/2017 and TUT Acoustic Scenes 2016 datasets indicate that the proposed method improves the performance of SED and ASC by 1.31 and 1.80 percentage points in terms of the F-score, respectively, compared with the conventional CRNN-based method.
\end{abstract}
%
\section{Introduction}
\label{sec:intro}
There has been renewed interest in the automatic analysis of various environmental sounds in our everyday lives \cite{Imoto_AST2018_01}.
The ability to automatically analyze environmental sounds will give rise to various applications, such as anomalous sound detection systems \cite{abnormal}, automatic life-logging systems \cite{Stork_ROMAN2012_01,Imoto_INTERSPEECH2013_01}, monitoring systems \cite{surveillance,Ntalampiras_ICASSP2009_01}, and hearing-impaired support systems \cite{healthcare,water}.
In environmental sound analysis, two tasks have primarily been studied: sound event detection (SED) and acoustic scene classification (ASC).
SED is the identification of sound event labels and detection of their boundaries in audio signals, where a sound event represents a type of sound such as ``car horn,'' ``people walking,'' or ``keyboard typing.''
On the other hand, ASC is the identification of acoustic scene labels from a recorded sound, where an acoustic scene represents a recording location, situation, or human activity such as ``city center,'' ``on the bus,'' or ``cooking.'' 
For SED and ASC, various methods using the Gaussian mixture model (GMM) \cite{SED_GMM,ASC_GMM}, hidden Markov model (HMM) \cite{SED_HMM,ASC_HMM}, and support vector machine (SVM) \cite{ASC_SVM}, have been proposed.
In SED, since multiple sound events often occur simultaneously, polyphonic SED methods, which can detect multiple overlapping sound events, have been proposed.
Some works have applied non-negative matrix factorization (NMF) \cite{SED_NMF,Komatsu_DCASE2016_01} for polyphonic SED, wherein the spectrogram of input sound is decomposed into a product of a basis and activation matrix.
Each basis and activation vector represent a single sound event and the active duration of the corresponding event, respectively.
More recently, in sound event and acoustic scene analysis there has been a growing trend towards the use of deep neural networks (DNNs) \cite{DCASE2017_SED,DCASE2017_ASC}, which outperform the conventional methods based on the GMM, HMM, and SVM.

\begin{figure*}[t!]
\centering
\includegraphics[width=1.90\columnwidth]{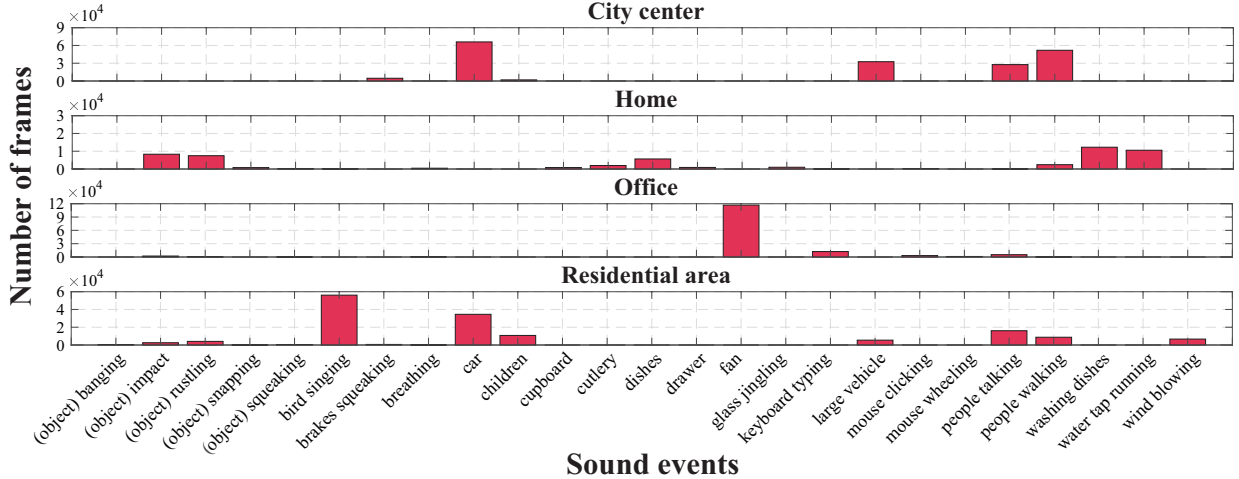}
\vspace{-8pt}
\caption{Number of frames of sound events for each acoustic scene in dataset used for our evaluation experiments.}
\label{fig:number_of_frames}
\vspace{5pt}
\end{figure*}

None of the studies on SED and ASC have tackled sound event and acoustic scene analysis jointly; however, in many cases, sound events and acoustic scenes are associated with each other.
For instance, in the acoustic scene ``office,'' the sound events ``mouse clicking'' and ``keyboard typing'' are likely to occur, whereas the sound events ``car'' and ``bird singing'' rarely occur.
That is, when analyzing the sound events ``mouse clicking'' and ``keyboard typing,'' it is expected that information on the acoustic scene ``office'' will help recognize these sound events.
Similarly, information on the sound events ``mouse clicking,'' and ``keyboard typing'' can help to predict the acoustic scene ``office.''
On the basis of this concept, SED taking information on acoustic scenes into account \cite{Mesaros_EUSIPCO2011_01,Heittola_JASM2013_01,Imoto_IEICE2016_01} and ASC utilizing information on sound events \cite{Imoto_TASLP2019_01} have been proposed.
However, these conventional methods do not analyze both sound events and scenes simultaneously.
Moreover, these methods cannot be applied to state-of-the-art neural-network-based methods.

To address this problem, we have proposed a method for joint analysis of sound events and acoustic scenes using multitask learning \cite{Tonami_WASPAA2019_01}.
Multitask learning is a method for learning related multiple tasks simultaneously by sharing the parameters of machine learning models \cite{Caruana1997}.
In speech processing, it has been reported that multitask learning improves the performance of speech recognition \cite{Giri_ICASSP2015_01,Kim_ICASSP2017_01}.
\textcolor{black}{
In the analysis of environmental sounds, ASC using environment-based scene grouping and multitask learning of multiple classification networks of grouped scenes (e.g., indoor/outdoor scene) \cite{ASC_MTL}, and SED with the multitask learning of event detection and timestamp identification \cite{SED_MTL01,SED_MTL02}, have been proposed.}
In this paper, we discuss the method for joint analysis of sound events and acoustic scenes using multitask learning in detail, and we investigate the performance of the proposed multitask-learning-based method in detail.
The contribution of the proposed method is as follows.

\vspace{5pt}
\begin{itemize}
  \setlength{\itemsep}{5pt}
  \item We discuss in detail a new method for joint analysis of sound events and acoustic scenes utilizing multitask learning combining SED and ASC methods.
  \item We demonstrate that the multitask-learning-based method improves SED and ASC performances.
\end{itemize}
\vspace{3pt}

The rest of this paper is organized as follows.
In Section 2, we discuss conventional methods for SED and ASC and the proposed method for the joint analysis of sound events and scenes utilizing multitask learning.
In Section 3, we report the results of experiments conducted to evaluate the performance of event detection and scene classification.
We then conclude this paper in Section 4.
%
\section{Conventional Methods for SED and ASC}
\label{ssec:conventional}
%
In this section, we overview conventional SED and ASC methods using neural networks.

SED involves the estimation of sound event categories and their time boundaries.
In recent years, a number of neural-network-based approaches, such as those using the convolutional neural network (CNN) \cite{Hershey_ICASSP2017_01}, recurrent neural network (RNN) \cite{Hayashi_TASLP2017_01}, and convolutional recurrent neural network \cite{SED_CRNN,Imoto_ICASSP2019_01}, have been proposed.
Specifically, it has been reported that methods in which the CNN and bidirectional gated recurrent unit (BiGRU) are combined detect sound events with reasonable performance \cite{SED_CRNN,Imoto_ICASSP2019_01}.
In the CNN--BiGRU-based method, the acoustic feature ${\bf X} \in \mathbb{R}^{D \times T}$, which is a time--frequency representation of the acoustic signal, is input to the network.
Here, $D$ and $T$ are the numbers of frequency bins and time frames of the input feature map, respectively.
The convolutional layer convolutes the input feature map with two-dimensional filters, and max pooling is then performed to reduce the dimension $D$ of the feature map.
The output of the convolutional layer ${\bf c}_{t}^{(d',c)}$ in time frame $t$ $\in \mathbb{R}^{D' \times C}$ is then concatenated as ${\bf c'} = ({\bf c}_{t}^{(1,1)}, {\bf c}_{t}^{(1,2)}, \ldots, {\bf c}_{t}^{(1,C)}, {\bf c}_{t}^{(2,1)}, \ldots, {\bf c}_{t}^{(D',C)}$), and then ${\bf c'}$ is fed to layers of the BiGRU, where $C$ is the number of convolution filters.
After that, the output vector ${\bf h}_{t}$ is calculated using the following set of equations:

%
\begin{align}
{\bf g}^{f}_{t} &= \sigma({\bf W}^{f}_{g} {\bf c}'_{t} + {\bf U}^{f}_{g} {\bf h}_{t-1} + {\bf b}^{f}_{g}),\\[1pt]
{\bf r}^{f}_{t} &= \sigma({\bf W}^{f}_{r} {\bf c}'_{t} + {\bf U}^{f}_{r} {\bf h}_{t-1} + {\bf b}^{f}_{r}),\\[1pt]
{\bf h}^{f}_{t} &= (1-{\bf g}^{f}_{t}) \odot {\bf h}_{t-1} \nonumber\\[-2pt]
&\hspace{10pt}+ {\bf g}^{f}_{t} \odot \tanh ({\bf W}^{f}_{h} {\bf c}'_{t} + {\bf U}^{f}_{h} ( {\bf r}^{f}_{h} {\bf h}_{t-1}) + {\bf b}^{f}_{h}),\\[1pt]
{\bf g}^{b}_{t} &= \sigma({\bf W}^{b}_{g} {\bf c}'_{t} + {\bf U}^{b}_{g} {\bf h}_{t+1} + {\bf b}^{b}_{g}),\\[1pt]
{\bf r}^{b}_{t} &= \sigma({\bf W}^{b}_{r} {\bf c}'_{t} + {\bf U}^{b}_{r} {\bf h}_{t+1} + {\bf b}^{b}_{r}),\\[1pt]
{\bf h}^{b}_{t} &= (1-{\bf g}^{b}_{t}) \odot {\bf h}_{t+1} \nonumber\\[-2pt]
&\hspace{10pt}+ {\bf g}^{b}_{t} \odot \tanh ({\bf W}^{b}_{h} {\bf c}'_{t} + {\bf U}^{b}_{h} ( {\bf r}^{b}_{h} {\bf h}_{t+1}) + {\bf b}^{b}_{h}),\\[1pt]
{\bf h}_{t} &=
\begin{bmatrix}
{\bf h}^{f}_{t}\\[3pt]
{\bf h}^{b}_{t}
\end{bmatrix},
\label{eq:bigru}
\end{align}
\vspace{8pt}
\begin{figure*}[t!]
\centering
\includegraphics[width=1.90\columnwidth]{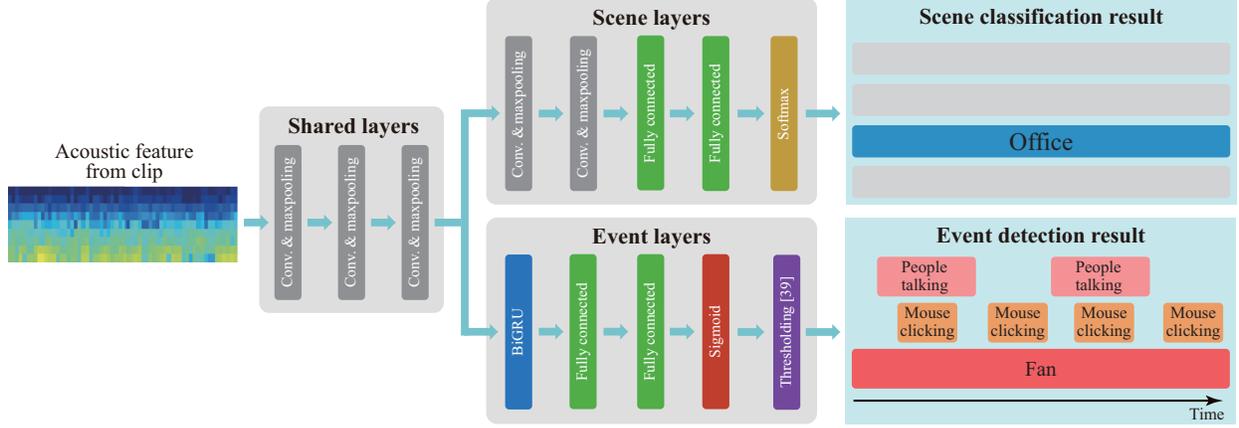}
\caption{Example of network structure for proposed multitask learning of sound events and acoustic scenes. \textcolor{black}{In shared layers, information on sound events  and scenes in common is extracted. In event and scene layers, event- and scene-specific features are extracted, respectively.}}
\label{fig:network}
\end{figure*}

\noindent where ${\bf W}$ and ${\bf U}$ are parameter matrices and ${\bf b}$ is a bias vector.
Subscripts $g$ and $r$ indicate the update gate and reset gate, respectively.
Superscripts $f$ and $b$ are the forward and backward networks, respectively.
${\bf g}$, ${\bf r}$, $\odot$, and $\sigma$ indicate the update gate vector, reset gate vector, Hadamard product, and sigmoid function, respectively.
The BiGRU layer is followed by a fully connected (FC) layer, and the output layer of the network is finally calculated as

\vspace{-5pt}
\begin{align}
{\bf y}_{t} = \sigma({\bf h}_{t}).
\label{eq:outputlayer_scene}
\end{align}
%

\noindent The network parameters of the CNN--BiGRU method are optimized with the following sigmoid cross-entropy objective function ${\mathcal L}_{\rm event}({\bi \Theta}_{1})$ using backpropagation through time~(BPTT):

\vspace{-5pt}
\begin{align}
\ {\mathcal L}_{\rm event} ({\bi \Theta}_{1}) &= - \! \sum^{T}_{t=1} {\big \{} {\bf z}_{t} \log ( {\bf y}_{t} ) + (1-{\bf z}_{t}) \log (1-{\bf y}_{t}) {\big \}} 
\nonumber\\[-1pt]
&\hspace{-35pt}= - \! \sum^{M}_{m=1} \sum^{T}_{t=1} \! {\Big \{} z_{m,t} \log y_{m,t} \! + \! (1 \! - \! z_{m,t}) \log {\big (} 1 \! - \! y_{m,t} {\big )} {\Big \}},
\label{eq:event_loss}
\end{align}

\noindent where ${\bi \Theta}_{1}$, $M$, and $m$ are the network parameters of the CNN--BiGRU method, the number of sound event categories, and the index of the event category, respectively.
$z_{m,t}$ is the target label in time frame $t$, which is 1 if sound event $m$ is active in time frame $t$, and 0 otherwise.

ASC involves the classification of an audio recording into one of the predefined classes of the acoustic scene.
Several lines of evidence indicate that CNN-based approaches achieve state-of-the-art performance in ASC \cite{Mesaros2017,Sakashita_DCASE2018_01}.
In CNN-based ASC, the time--frequency representation of acoustic signal ${\bf X}$ is fed to a two-dimensional convolutional layer, where the input feature map is also compressed, as with the CNN layer in SED.
The output of the CNN layer is then concatenated and input to a fully connected layer, and the output layer of the network is finally calculated as

\vspace{-5pt}
\begin{align}
{\bf y}_{t} &= S({\bf h}_{t}),
\label{eq:outputlayer_event}
\end{align}
%

\noindent where $S$ denotes the softmax function.
The network is optimized under the following softmax cross-entropy objective function 
${\mathcal L}_{\rm scene}({\bi \Theta}_{2})$:

\vspace{-5pt}
\begin{align}
{\mathcal L}_{\rm scene} ({\bi \Theta}_{2}) = - \sum^{N}_{n=1} {\Big \{} z_{n} \log S (y_{n}) {\Big \}},
\label{eq:scene_loss}
\end{align}

\noindent where ${\bi \Theta}_{2}$ is the parameter of the network for ASC.
$N$, $n$, and $z_n$ are the number of the acoustic scene categories, the index of the scene categories, and the target scene label, respectively.
%
%
%
\section{Joint Analysis of Sound Events and Scenes Based on Multitask Learning}
In conventional works addressing SED and ASC, they have been studied separately.
However, many sound events and acoustic scenes are related to each other.
Fig.~\ref{fig:number_of_frames} shows the number of time frames of the sound events for each acoustic scene in a development dataset used for experiments.
This figure indicates that the occurrence of sound events is biased for acoustic scenes; thus, sound events and acoustic scenes are key information in their mutual identification.
On the basis of this idea, we have proposed the multitask learning of sound events and acoustic scenes, in which the parts of the networks holding information on sound events and scenes in common are shared.
Multitask learning to jointly analyze multiple tasks has been proposed by Caruana \cite{Caruana1997}, and it has been reported that when multiple tasks are related to each other, it achieves better performance than a single-task method.

The concept of the proposed method is illustrated in Fig.~\ref{fig:network}.
We refer to the shared part of the network as the ``shared layers,'' and the following scene-specific layers and event-specific layers as the ``scene layers'' and ``event layers,'' respectively. 
\textcolor{black}{
The ``shared layers'' are expected to capture the information common to sound events and scenes from audio clips.
The features are fed into each specific layer to tune each model of SED/ASC.
The ``event'' and ``scene layers'' extract event- and scene-specific features, respectively.
After that, each output of specific layers is utilized for scene classification or event detection.
}
In this work, we apply the CRNN as the event detection network and the CNN as the scene classification network, and the layers of the CNN are shared between the event detection and scene classification networks.

To optimize this network, the following objective function is applied:

\vspace{-5pt}
\begin{align}
{\mathcal L}({\bi \Theta}) = \alpha {\mathcal L}_{\rm event}({\bi \Theta}_{1}) + \beta {\mathcal L}_{\rm scene}({\bi \Theta}_{2}),
\label{eq:loss}
\end{align}

\noindent where ${\bi \Theta}$ denotes the parameters of the proposed network.
$\alpha$ and $\beta$ are the weights of the objective functions for SED and ASC, respectively.
In this work, we set $\alpha$ as 1.0 because $\alpha$ can be set as 1.0 without loss of generality in many of optimization algorithms such as stochastic gradient descent (SGD) and Adam.
Note that the proposed method can also be applied to any network that can be optimized by Eq.~(\ref{eq:loss}).
%
%
%
\section{Experiments}
\subsection{Experimental Conditions}
%
To evaluate the performances of SED and ASC using the proposed multitask method, we conducted evaluation experiments using the TUT Sound Events 2016 \cite{Mesaros2016TUTDF}, TUT Sound Events 2017 \cite{Mesaros2017}, TUT Acoustic Scenes 2016 \cite{Mesaros2016TUTDF}, and TUT Acoustic Scenes 2017 \cite{Mesaros2017} datasets.
From these datasets, we selected sound clips including four acoustic scenes, ``home,'' ``residential area'' (TUT Sound Events 2016), ``city center'' (TUT Sound Events 2017, TUT Acoustic Scenes 2017), and ``office'' (TUT Acoustic Scenes 2016), which contain 266 min (development set: 192 min, evaluation set: 74 min) of audio.
Here, the acoustic scene ``office'' in TUT Acoustic Scenes 2016 and ``city center'' in TUT Acoustic Scenes 2017 did not have sound event labels; thus, we manually annotated the sound clips with sound event labels by the procedure described in \cite{Mesaros2016TUTDF,Mesaros2017}.
These sound clips include the 25 types of sound event labels listed in Fig.~\ref{fig:number_of_frames}.
The sound event labels annotated for this experiment are available in \cite{Imoto_dataset2019_01}.

As acoustic features, we used 64-dimensional log-mel energies calculated for each 40 ms time frame with 50\% overlap.
This setting is based on the baseline system of DCASE2018 Challenge task4 \cite{setting01} and a work that won the first place in DCASE2019 Challenge task4 \cite{setting02}.
The acoustic features were then input to the network shown in Fig.~\ref{fig:network}.
The model parameter was tuned using only the development dataset.
We then evaluated the performance of our proposed model using evaluation dataset.
Active sound events were predicted using adaptive thresholds \cite{Xu_DCASE2017_adaptive_threshold} tuned for each sound event using the development dataset.
Each model was trained and evaluated ten times with various initial values of model parameters.
Other experimental conditions are listed in Table~\ref{tbl:parameter}. 

\textcolor{black}{
As comparative methods, we also evaluated the detection performance for sound events by CNN method (referred to as CNN (event)) and the  CNN--BiGRU method (referred to as CRNN (event)), and the classification performance for acoustic scenes by the CNN method (referred to as CNN (scene)).
CRNN (event) and CNN (event) have the same structures as the shared + event and shared layers in Fig.~\ref{fig:network}, respectively.
CNN (scene) has the same structure as the shared and scene layers in the proposed method.
Moreover, to verify the utility of the proposed method, we compared the proposed method with a SED method based on multitask learning of SED and sound activity detection \cite{SED_MTL01} (referred to as SED + SAD), and the proposed method + the conventional method using SAD \cite{SED_MTL01} (referred to as proposed + SAD).
}
\begin{table}[t]
\small
\vspace{-5pt}
\caption{Experimental conditions}
\label{tbl:parameter}
\centering
\begin{tabular}{ll}
\wcline{1-2}
&\\[-9pt]
Acoustic feature & Log-mel energy (64 dim.)\\
Frame length \hspace{-3pt} / \hspace{-3pt} shift & 40 ms \hspace{-3pt} / \hspace{-3pt} 20 ms\\
Length of sound clip & 10 s\\\hline
&\\[-10pt]
{\bf Shared layer}& \\
\ \ \ Network of shared layers & 3 CNN\\
\ \ \ \# channels of CNN layers & 128, 128, 128 \\
\ \ \ Filter size & 3$\times$3 \\
\ \ \ Pooling size & 8$\times$1, 2$\times$1, 2$\times$1 (max pooling) \\\hline
&\\[-10pt]
{\bf Scene layer}& \\
\ \ \ Network of scene layers & 2 CNN \& 1 FC layers\\
\ \ \ \# channels of CNN layers & 256, 256 \\
\ \ \ Filter size & 3$\times$3 \\
\ \ \ Pooling size & 1$\times$25, 1$\times$20 (max pooling) \\
\ \ \ \# units in FC layer & 32 \\\hline
&\\[-10pt]
{\bf Event layer}& \\
\ \ \ Network of event layers & 1 BiGRU \& 1 FC layers \\
\ \ \ \# units in GRU layer & 32 \\
\ \ \ \# units in FC layer & 32 \\
\wcline{1-2}
\end{tabular}
\vspace{0pt}
\end{table}
%
%
%
\vspace{-3pt}
\subsection{Metrics}
In SED, since sound events may overlap, the event detection performance was evaluated on the basis of a segment-based F-score and error rate (ER) \cite{Mesaros2016_MDPI}.
To calculate the segment-based F-score, the precision and recall were first calculated as

\vspace{-8pt}
\begin{align}
\label{eq:Precision}
{\rm Precision} &= \frac{{\rm TP}}{{\rm TP}+{\rm FP}},\\[0pt]
{\rm Recall} &= \frac{{\rm TP}}{{\rm TP}+{\rm FN}},
\label{eq:Recall}
\end{align}
\vspace{-5pt}

\noindent where TP, FP, and FN are the total counts of true positive, false positive, and false negative for all time frames and sound events, respectively.
The segment-based F-score was then calculated as 

\begin{align}
{\rm F\textrm{-}score} = \frac{2 \cdot {\rm Precision} \cdot {\rm Recall}}{{\rm Precision} + {\rm Recall}}.
\label{eq:F}
\end{align}

\noindent To calculate the segment-based ER, we first calculated substitutions (S), deletions (D), and insertions (I) as

\begin{align}
{\rm S}(k) &= \min({\rm FN}(k), {\rm FP}(k)),\\[3pt]
{\rm D}(k) &= \max(0, {\rm FN}(k)-{\rm FP}(k)),\\[3pt]
{\rm I}(k) &= \max(0, {\rm FP}(k)-{\rm FN}(k)),
\label{eq:SDI}
\end{align}

\noindent where $k$ indicates the index of the time frame.
We then calculated the ER as

\begin{align}
{\rm ER} =  \frac{\sum_{k=1}^{K} {\rm S}(k) + \sum_{k=1}^{K} {\rm D}(k) + \sum_{k=1}^{K} {\rm I}(k)}{\sum_{k=1}^{K} {\rm N}(k)},
\label{eq:ER}
\end{align}

\noindent where $K$ and $N(k)$ are the total number of time frames and the number of sound events in time frame $k$, respectively.

On the other hand, ASC is a simple classification task and acoustic scenes do not overlap in time.
We thus calculated the F-score using Eqs.~(\ref{eq:Precision}) -- (\ref{eq:F}), where TP, FP, and FN are the total numbers of sound clips of true positive, false positive, and false negative, respectively.

\begin{table}[t]
\small
\caption{\textcolor{black}{Performances of sound event detection and scene classification}}
\vspace{1pt}
\label{tbl:F-score}
\centering
\begin{tabular}{cccc}
\wcline{1-4}
&&&\\[-8pt]
\multirow{2}{*}{Method} & \multicolumn{2}{c}{Event} & Scene \\\cline{2-4}
&&&\\[-9pt]
& F-score & ER  & F-score \\
\wcline{1-4}
&&&\\[-7pt]
\multicolumn{1}{l}{\textcolor{black}{CNN (event)}} & \textcolor{black}{34.67\%} & \textcolor{black}{1.310} & -\\
\multicolumn{1}{l}{CRNN (event)} & 44.88\% & 1.301 & -\\
\multicolumn{1}{l}{CNN (scene)} & - & - & 67.35\% \\
\multicolumn{1}{l}{\textcolor{black}{SED + SAD \cite{SED_MTL01}}} & \textcolor{black}{47.59\%} & \textcolor{black}{1.081} & -  \\\hline
&&&\\[-8pt]
\multicolumn{1}{l}{Proposed ($\beta$=10.0)} & 21.85\%& 4.531 & {\bf 69.15\%} \\
&&&\\[-10pt]
\multicolumn{1}{l}{Proposed ($\beta$=0.01)} & 45.76\%& 1.119 & 65.17\% \\
\multicolumn{1}{l}{Proposed ($\beta$=0.0001)} & {\bf 46.19\%} & {\bf 1.102} & 61.89\% \\
\multicolumn{1}{l}{\textcolor{black}{Proposed + SAD ($\beta=0.01$)}} & \textcolor{black}{\bf 48.99\%} & \textcolor{black}{\bf 0.908} & \textcolor{black}{60.02\%} \\ \wcline{1-4}
\end{tabular}
\vspace{5pt}
\end{table}
\begin{figure}[t!]
\centering
\includegraphics[width=0.94\columnwidth]{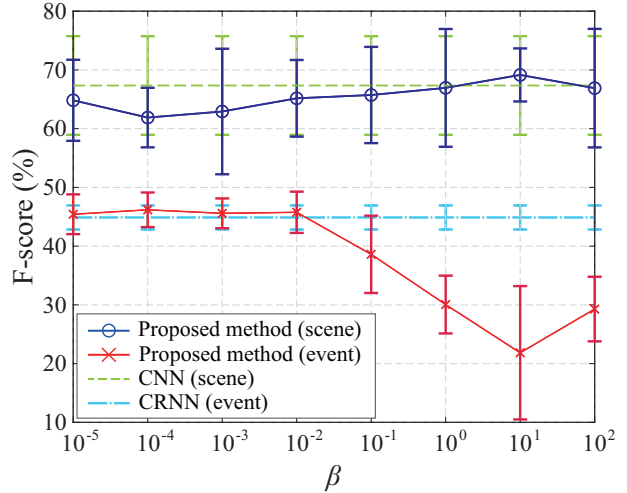}
\vspace{-8pt}
\caption{\textcolor{black}{Sound event detection and scene classification performance as functions of weight $\beta$}}
\label{fig:beta_relation}
\end{figure}
\begin{table*}[t]
\scriptsize
\caption{Sound event detection performance for each event}
\centering
\scalebox{1.013}[1.013]{
\hspace{-3.2pt}
\begin{tabular}{p{12mm}lp{7.8mm}p{7.8mm}p{7.8mm}p{7.8mm}p{7.8mm}p{7.8mm}p{7.8mm}p{7.8mm}p{7.8mm}p{7.8mm}p{7.8mm}p{7.8mm}}
\wcline{1-14}
&&&&&&&&&&&&&\\[-6pt]
\multicolumn{2}{c}{\multirow{2}{*}{Event}} & \multicolumn{1}{c}{(object)} & \multicolumn{1}{c}{(object)} & \multicolumn{1}{c}{(object)} & \multicolumn{1}{c}{(object)} & \multicolumn{1}{c}{(object)} & \multicolumn{1}{c}{bird} & \multicolumn{1}{c}{brakes} &\multicolumn{1}{c}{\multirow{2}{*}{breathing}}&\multicolumn{1}{c}{\multirow{2}{*}{car}}&\multicolumn{1}{c}{\multirow{2}{*}{children}}& \multicolumn{1}{c}{\multirow{2}{*}{cupboard}}&\multicolumn{1}{c}{\multirow{2}{*}{cutlery}}\\[-1pt]
&& banging & impact & rustling & snapping & squeaking & singing & squeaking &&&&&\\[0pt]
\wcline{1-14}
&&&&&&&&&&&&&\\[-5.5pt]
CRNN & F-score & 0.00\%  & \textbf{9.56\%} & \textbf{13.89\%} & 0.00\% & 0.00\% & 47.01\% & \textbf{7.73\%} & 0.00\% & \textbf{51.75\%} & \textbf{6.10\%} & 0.00\% & \textbf{0.39\%} \\\cline{2-14}
(event)&&&&&&&&&&&&&\\[-7pt]
  & ER & \textbf{1.007} & 1.346 & 3.542 & \textbf{1.002} & 1.000 & 1.538 & 1.022 & \textbf{1.008} & 1.740 & 1.491 & 1.000 & 1.018\\\hline
\\[-7pt]
\\[-7pt]
Proposed& F-score & 0.00\% & 8.82\% & 13.30\% & 0.00\% & 0.00\% & \textbf{48.88\%} & 6.61\% & 0.00\% & 50.97\% & 3.54\% & 0.00\% & 0.11\% \\\cline{2-14}
($\beta$=0.0001)&&&&&&&&&&&&&\\[-7pt]
 & ER & 1.122 & \textbf{1.252} & \textbf{3.393} & 1.004 & 1.000 & \textbf{1.448} & \textbf{0.975} & 1.012 & \textbf{1.686} & \textbf{1.334} & 1.000 & \textbf{1.004}\\
\wcline{1-14}\\[-13pt]
\end{tabular}
}
\hspace*{-3pt}
\begin{tabular}{p{12.2mm}p{7.8mm}p{7mm}p{7mm}p{7mm}p{7mm}p{7mm}p{7mm}p{7mm}p{7mm}p{7mm}p{7mm}p{7mm}p{7mm}p{7mm}}
\\[5pt]
\wcline{1-15}
&&&&&&&&&&&&&&\\[-6pt]
\multicolumn{2}{c}{\multirow{2}{*}{Event}}&\multicolumn{1}{c}{\multirow{2}{*}{\!dishes\!}}& \multicolumn{1}{c}{\multirow{2}{*}{\!drawer\!}} & \multicolumn{1}{c}{\multirow{2}{*}{\!fan\!}} & \multicolumn{1}{c}{\!glass\!} & \multicolumn{1}{c}{\!keyboard\!} & \multicolumn{1}{c}{\!large\!} & \multicolumn{1}{c}{\!mouse\!} & \multicolumn{1}{c}{\!mouse\!} & \multicolumn{1}{c}{\!people\!} & \multicolumn{1}{c}{\!people\!} & \multicolumn{1}{c}{\!washing\!} & \multicolumn{1}{c}{\!water tap\!} & \multicolumn{1}{c}{\!wind\!} 
\\[-1pt]
&\!\!&\!\!&\!\!&\!\!&\multicolumn{1}{c}{\!jingling\!}&\multicolumn{1}{c}{\!typing\!}&\multicolumn{1}{c}{\!vehicle\!}&\multicolumn{1}{c}{\!clicking\!}&\multicolumn{1}{c}{\!wheeling\!}&\multicolumn{1}{c}{\!talking\!}&\multicolumn{1}{c}{\!walking\!}&\multicolumn{1}{c}{\!dishes\!}&\multicolumn{1}{c}{\!running\!}&\multicolumn{1}{c}{\!blowing\!}\\[0pt]
\wcline{1-15}
&&&&&&&&&&&&&\\[-5.5pt]
CRNN & F-score & \textbf{11.29\%} & 0.00\% & 71.97\% & 0.00\% & 4.29\% & 17.55\% & 0.02\% & 0.00\% & 2.48\% & \textbf{15.34\%} & \textbf{38.44\%} & \textbf{66.32\%} & \textbf{15.96\%} \\\cline{2-15}
(event)&&&&&&&&&&&&&&\\[-7pt]
& ER & 1.076 & 1.000 & 0.733 & 1.011 & 1.805 & 4.642 & 1.025 & 1.003 & 3.391 & 1.559 & 2.029 & \textbf{0.659} & 1.036 \\\hline
\\[-7pt]
Proposed & F-score & 6.94\% & 0.00\% & \textbf{72.31\%} & 0.00\% & \textbf{5.00\%} & \textbf{20.12\%} & \textbf{0.30\%} & 0.00\% & \textbf{2.75\%} & 15.04\% & 35.54\% & 60.80\% & 8.81\% \\\cline{2-15}
($\beta$=0.0001)&&&&&&&&&&&&&&\\[-7pt]
 & ER & \textbf{1.041} & 1.000 & \textbf{0.633} & \textbf{1.005} & \textbf{1.389} & \textbf{2.634} & \textbf{1.007} & \textbf{1.001} & \textbf{2.725} & \textbf{1.396} & \textbf{1.845} & 0.707 & \textbf{0.997} \\
\wcline{1-15}
\end{tabular}
\label{tbl:each_event}
\vspace{0pt}
\end{table*}
%
%
%
%
\subsection{Experimental Results}
\subsubsection{Overall Performances of SED and ASC}
Table~\ref{tbl:F-score} shows the experimental results of using the conventional methods and the proposed multitask-based method.
The results show that the proposed multitask-based method enables the joint analysis of sound events and acoustic scenes with a reasonable performance compared with CRNN (event) and CNN (scene).
Moreover, when $\beta$ = 0.0001 (Proposed), the F-score of event detection by the proposed method was improved by 1.31 points compared with that of the conventional CRNN (event).
When $\beta$ = 10.0 (Proposed), the F-score of scene classification of the proposed method was improved by 1.80 percentage points compared with that of the conventional CNN (scene).
These results indicate that information on sound events and acoustic scenes are mutually helpful in SED method and ASC.

\textcolor{black}{To further demonstrate the utility of the proposed method, we applied our method to the conventional method for SED \cite{SED_MTL01}, which is based on the multitask learning of SED and SAD.
The experimental results show that the event detection performance of SED + SAD \cite{SED_MTL01} is better than that of CRNN (event). 
Moreover, the F-score of event detection in proposed + SAD ($\beta$= 0.01) was improved by 1.4 percentage points compared with that of SED + SAD \cite{SED_MTL01}. 
The results indicate that the proposed method improves the SED performance when using not only CNN--BiGRU but also other models.}

Further experiments were conducted to examine how the proposed method performs with various $\beta$.
Fig.~\ref{fig:beta_relation} shows the performances of event detection and scene classification for various $\beta$.
The results show that when $\beta$ is \textcolor{black}{less than or equal to 0.01,} the SED performance of the proposed method is better than that of CRNN (event).
When $\beta$ is larger than \textcolor{black}{1}, the proposed method achieves a better ASC performance than CNN (scene), whereas the detection performance of sound events by the proposed method was worse than that by the conventional CRNN (event).
When $\beta$ takes a large or small value, the ASC and SED performances of the proposed method agree well with those of the conventional CNN (scene) and CRNN (event), respectively.
This is because when $\beta$ is small, the proposed method approximately works as CRNN (event), whereas when $\beta$ is large, the proposed method tends to ignore the sound event loss ${\bi \Theta}_{1}$.
\begin{figure}[t!]
\vspace{5pt}
\centering
\subfigure{%
\includegraphics[scale=0.68]{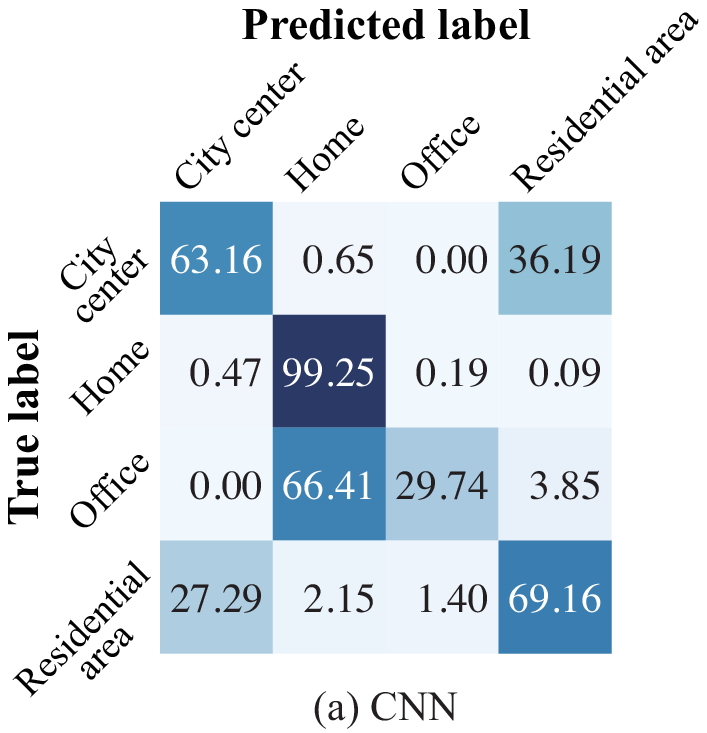}\vspace{-10pt}}%
\subfigure{%
\includegraphics[scale=0.68]{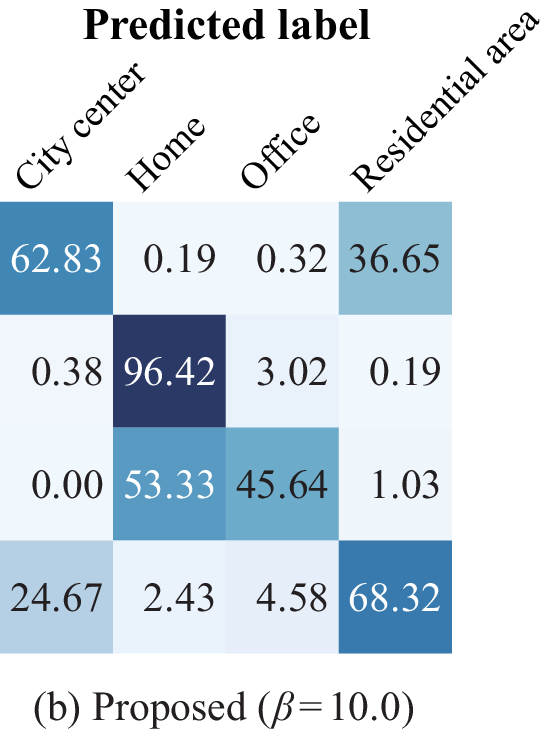}}%
\caption{Performance of acoustic scene classification for each scene in terms of recall (\%)}
\label{fig:ASC_performance}
\vspace{5pt}
\end{figure}
\begin{table*}[t]
\scriptsize
\caption{Performance of SED for each event and scene in terms of false-positive rate (FPR)}
\label{tbl:each_scene_event}
\centering
\begin{tabular}{llcccccccccccc}
\wcline{1-14}
&&&&&&&&&&&&&\\[-6pt]
\multicolumn{2}{c}{\multirow{2}{*}{Event}}&\multicolumn{1}{c}{(object)}&\multicolumn{1}{c}{(object)}&\multicolumn{1}{c}{(object)}&\multicolumn{1}{c}{(object)}&\multicolumn{1}{c}{(object)}&\multicolumn{1}{c}{bird}&\multicolumn{1}{c}{brakes}&\multicolumn{1}{c}{\multirow{2}{*}{breathing}}&\multicolumn{1}{c}{\multirow{2}{*}{car}}&\multicolumn{1}{c}{\multirow{2}{*}{children}}&\multicolumn{1}{c}{\multirow{2}{*}{cupboard}}&\multicolumn{1}{c}{\multirow{2}{*}{cutlery}}
\\[-1pt]
&&\multicolumn{1}{c}{banging}&\multicolumn{1}{c}{impact}&\multicolumn{1}{c}{rustling}&\multicolumn{1}{c}{snapping}&\multicolumn{1}{c}{squeaking}&\multicolumn{1}{c}{singing}&\multicolumn{1}{c}{squeaking}&&&&&
\\\wcline{1-14}\\[-5pt]
City& CRNN &\textbf{0.0000}&0.0070&0.0001&0.0000&0.0000&0.1150&0.0004&0.0000&0.5835&0.0043&0.0000&0.0000\\\cline{2-14}
center&&&&&&&&&&&&&\\[-5pt]
&Proposed &0.0002&\textbf{0.0014}&\textbf{0.0000}&0.0000&0.0000&\textbf{0.1073}&\textbf{0.0001}&0.0000&\textbf{0.5457}&\textbf{0.0038}&0.0000&0.0000\\[0pt]\hline
\\[-5pt]
\multirow{2}{*}{Home}& CRNN &0.0000&0.0409&0.1175&0.0000&0.0000&0.0146&0.0002&\textbf{0.0000}&\textbf{0.0263}&0.0041&0.0000&0.0004\\\cline{2-14}
&&&&&&&&&&&&&\\[-5pt]
& Proposed &0.0000&\textbf{0.0352}&\textbf{0.1110}&0.0000&0.0000&\textbf{0.0115}&\textbf{0.0000}&0.0001&0.0367&\textbf{0.0010}&0.0000&\textbf{0.0001}\\[0pt]\hline
\\[-5pt]
\multirow{2}{*}{Office}&CRNN&0.0000&\textbf{0.0011}&\textbf{0.0041}&0.0000&0.0000&0.0062&0.0000&0.0000&0.0000&\textbf{0.0002}&0.0000&0.0000\\\cline{2-14}
&&&&&&&&&&&&&\\[-5pt]
& Proposed &0.0000&0.0036&0.0073&0.0000&0.0000&\textbf{0.0028}&0.0000&0.0000&0.0000&0.0003&0.0000&0.0000\\[0pt]\hline
\\[-5pt]
Residential& CRNN & 0.0000&0.0026&0.0094&0.0000&0.0000&0.3011&0.0038&0.0000& 0.1837&0.0049&0.0000&0.0000\\\cline{2-14}
area &&&&&&&&&&&&&\\[-5pt]
& Proposed & 0.0000& \textbf{0.0008}&\textbf{0.0056}&0.0000&0.0000&\textbf{0.2696}&\textbf{0.0005}&0.0000&\textbf{0.1684}&\textbf{0.0036}&0.0000&0.0000\\[0pt]
\wcline{1-14}
\end{tabular}
\ \\[6pt]
\hspace*{0.3pt}
\begin{tabular}{llccccccccccccc}
\wcline{1-15}
&&&&&&&&&&&&&&\\[-5.5pt]
\multicolumn{2}{c}{\multirow{2}{*}{Event}}\!&\!\multirow{2}{*}{dishes}\!&\!\multirow{2}{*}{drawer}\!&\!\multirow{2}{*}{fan}\!&\!glass\!&\!keyboard\!&\!large\!&\!mouse\!&\!mouse\!&\!people\!&\!people\!&\!washing\!&\!water tap\!&\!wind
\\[0pt]
\!&\!\!&\!\!&\!\!&\!\!&\!jingling\!&\!typing\!&\!vehicle\!&\!clicking\!&\!wheeling\!&\!talking\!&\!walking\!&\!dishes\!&\!running\!&\!blowing\\
&&&&&&&&&&&&&&\\[-15pt]
\\\wcline{1-15}
\\[-5pt]
City& CRNN &0.0007&0.0000&\textbf{0.0024}&0.0000&0.0007&0.3268&0.0000&0.0000&0.0521&0.0656&0.0000&0.0009&\textbf{0.0000}\\\cline{2-15}
center&&&&&&&&&&&&&&\\[-5pt]
& Proposed &0.0007&0.0000&0.0041&0.0000&\textbf{0.0000}&\textbf{0.1552}&0.0000&0.0000&\textbf{0.0342}&\textbf{0.0434}&0.0000&\textbf{0.0004}&0.0002\\[0pt]\hline
\\[-5pt]
\multirow{2}{*}{{Home}}& CRNN &0.0085&0.0000&0.4151&0.0001&0.0304&0.0002&0.0004&0.0000&0.0242&0.0008&0.2034&0.0815&0.0010\\\cline{2-15}
&&&&&&&&&&&&&&\\[-5pt]
& Proposed &\textbf{0.0042}&0.0000&\textbf{0.2816}&0.0000&\textbf{0.0147}&\textbf{0.0000}&\textbf{0.0001}&0.0000&\textbf{0.0075}&\textbf{0.0003}&\textbf{0.1663}&\textbf{0.0680}&\textbf{0.0005}\\[0pt]\hline
\\[-5pt]
\multirow{2}{*}{Office}& CRNN &0.0000&0.0000&0.0011&0.0000&0.0052&0.0000&0.0000&0.0000&0.0061&0.0001&0.0023&0.0000&0.0000\\\cline{2-15}
&&&&&&&&&&&&&&\\[-5pt]
& Proposed &0.0000&0.0000&\textbf{0.0008}&0.0000&\textbf{0.0039}&0.0000&0.0000&0.0000&\textbf{0.0020}&0.0001&\textbf{0.0009}&0.0000&0.0000\\[0pt]\hline
\\[-5pt]
Residential& CRNN &0.0000&0.0000&0.0531&0.0000&0.0003&0.0099&0.0000&0.0000&0.1124&0.0697&0.0000&0.0009&0.0041\\\cline{2-15}
area&&&&&&&&&&&&&&\\[-5pt]
& Proposed &0.0000&0.0000&\textbf{0.0480}&0.0000&\textbf{0.0002}&\textbf{0.0029}&0.0000&0.0000&\textbf{0.0983}&\textbf{0.0589}&0.0000&\textbf{0.0001}&\textbf{0.0010}\\[0pt]
\wcline{1-15}
\end{tabular}
\label{tbl:each_event_scene}
\vspace{5pt}
\end{table*}
%
%
%
%
%
%
%
%
%
\subsubsection{Detailed Investigation of ASC Performance}
Fig.~\ref{fig:ASC_performance} shows confusion matrices of the scene classification performances in terms of recall.
The left and right confusion matrices depict the performances of the conventional CNN (scene) and the proposed multitask-based method ($\beta$=10.0), respectively. 
The results indicate that the proposed method achieves better classification performance than the conventional CNN (scene).
In particular, the recall for the acoustic scene ``office'' increases by 15.90 percentage points compared with that of the conventional CNN (scene).
This is because the acoustic scene ``office'' has the dominant sound event ``fan,'' which occurs only in the acoustic scene ``office,'' and thus, information on the sound event ``fan'' significantly contributes to the classification of the acoustic scene ``office.''
On the other hand, little improvement of the scene classification performance for the acoustic scenes ``city center'' and ``residential area'' was observed.
In the acoustic scenes ``city center'' and ``residential area,'' common sound events ``car,'' ``children,'' and ``large vehicle'' occur; thus, information of sound events may make a smaller contribution to the classification of these acoustic scenes.
%
%
\subsubsection{Sound Event Detection Performance for Each Sound Event and Scene}
%
We list the detection results of sound events for each event in Table~\ref{tbl:each_event}.
In this experiment, we evaluated the detection performance using $\beta = 0.0001$.
The proposed method achieves a better ER than the conventional CRNN (event) in most of the sound events.
These results show that the proposed method reduces the misdetection of sound events.
Moreover, the sound events ``fan,'' ``keyboard typing,'' and ``mouse clicking'' can be detected more accurately by the proposed method.
As shown in Fig.~\ref{fig:number_of_frames}, this is because these events are closely related to the particular acoustic scene ``office,'' and such information on the acoustic scene is helpful in detecting the sound events.

To investigate how the misdetection of sound events is reduced in the proposed method, we show the misdetection rate of the sound events for each event and scene in Table~\ref{tbl:each_scene_event}.
In this experiment, we evaluated the detection performance using $\beta = 0.0001$.
To evaluate the rate of misdetection of sound events for each event and scene, we define the false-positive rate (FPR) because Eq.~(\ref{eq:ER}) cannot be calculated when $\sum_{k=1}^{K} {\rm N}(k)$ equals 0.
Then the FPR is calculated as

\begin{align}
{\rm FPR} =  \frac{\sum_{k=1}^{K} {\rm FP}(k)}{{\it K}}.
\label{eq:FPR}
\end{align}

\noindent The results show that the proposed multitask-based method achieves a much better FPR in most of the sound events than the conventional CRNN-based method.
In particular, the misdetection rate of the sound events ``bird singing'' and ``fan,'' which are closely related to the scenes ``residential area'' and ``city,'' respectively, can be significantly decreased by using the multitask-based method.

\subsubsection{\textcolor{black}{Calculation cost of models}}
\textcolor{black}{
To verify the calculation cost of our method, we compared the numbers of parameters of CRNN (event), CNN (scene), and the multitask-based model.
Table \ref{tbl:num_parameters} shows the number of  parameters of each model.
The results indicate that the multitask-based model has more parameters compared with CRNN (event).
On the other hand, CNN (scene) has a comparable number of parameters to the multitask-based model.  
This is because scene-specific layers are mainly composed of CNN structures with many channels.
}

\begin{table}[t]
\small
\caption{\textcolor{black}{Numbers of model parameters}}
\vspace{-10pt}
\label{tbl:num_parameters}
\centering
\begin{tabular}{lrrr}
&&&\\\wcline{1-4}
&&&\\[-7pt]
\multicolumn{1}{c}{Method} &\multicolumn{3}{c}{\# of parameters}\\\wcline{1-4}
&&&\\[-7pt]
CRNN (event) & & 355,801 &\\
&&&\\[-10pt]
CNN (scene) & & 1,200,036 & \\
Proposed & & 1,258,621 &\\
\wcline{1-4}
\end{tabular}
\end{table}
%
%

%
%
\section{Conclusion}
In this paper, we discussed in detail a proposed method for joint analysis of sound events and acoustic scenes using multitask learning.
In the proposed method, we applied a CRNN for SED and a CNN for ASC, and the CNN layers in both networks were shared in order to hold information on sound events and scenes in common. 
We integrated the objective functions of the event detection network and scene classification network using a weight parameter and optimized the network.
Experimental results indicated that the proposed method enables joint analysis of sound events and scenes with reasonable performance compared with the conventional CRNN- and CNN-based methods.
Moreover, the F-scores of the proposed method for SED and ASC are improved by 1.31 and 1.80 percentage point, respectively, compared with those of the conventional methods.
%
\section{Acknowledgement}
\label{sec:ack}
This work was supported by JSPS KAKENHI Grant Number JP19K20304 and the NVIDIA GPU Grant Program.
%
%
\bibliographystyle{ieicetr}
\bibliography{IEEEabrv,IEICE2019}

%
\end{document}